\documentclass[prl,twocolumn,letterpaper,showpacs,groupedaddress,superscriptaddress,nofootinbib,floatfix,preprintnumbers,tightenlines]{revtex4-1}

\usepackage[colorlinks,urlcolor=blue]{hyperref}
\usepackage{amsmath,amssymb,amsthm,amsfonts,bbold}
\usepackage{graphicx}
\usepackage{bm}
\usepackage{comment}
\usepackage{color}
\usepackage[section]{placeins}

\allowdisplaybreaks
\let\Oldsection\section
\renewcommand{\section}{\FloatBarrier\Oldsection}
\let\Oldsubsection\subsection
\renewcommand{\subsection}{\FloatBarrier\Oldsubsection}

\newcount\hour \newcount\hourminute \newcount\minute
\hour=\time \divide \hour by 60
\hourminute=\hour \multiply \hourminute by 60
\minute=\time \advance \minute by -\hourminute
\newcommand{\mydate}{\ \today \ - \number\hour :\number\minute}

\def\OMIT#1{{}}

\newcommand{\bit}[1]{\mbox{\boldmath$#1$}}

\newcommand{\BE}{\begin{equation}}
\newcommand{\EE}{\end{equation}}
\newcommand{\BA}{\begin{eqnarray}}
\newcommand{\EA}{\end{eqnarray}}

\def\OMIT#1{{}}
\def\nc{{N_c}}

\def\nc{{N_c}}

\def\yo2{{f_\pi^2}}
\def\llra{{\relbar\joinrel\longrightarrow}}
\def\mapright#1{{\smash{\mathop{\llra}\limits_{#1}}}}

\def\oneht{\textstyle{1\over 2} }

\def\onett{\textstyle{1\over 3} }

\def\threeht{\textstyle{3\over 2} }

\begin{document}

\title{Chiral symmetry breaking, entanglement, and the nucleon spin decomposition}

\preprint{NT@UW-19-06}

\author{Silas R.~Beane}

\author{Peter Ehlers}
\affiliation{Department of Physics, University of Washington, Seattle, WA 98195-1560, USA}

\date{\mydate}

\begin{abstract}
  \noindent 

The nucleon is naturally viewed as a bipartite system of valence spin
\textemdash defined by its non-vanishing chiral charge \textemdash and non-valence or sea
spin.  The sea spin can be traced over to give a reduced density
matrix, and it is shown that the resulting entanglement entropy acts
as an order parameter of chiral symmetry breaking in the nucleon. In
the large-$\nc$ limit, the entanglement entropy vanishes and the
valence spin accounts for all of the nucleon spin, while in the limit
of maximal entanglement entropy, the nucleon loses memory of the
valence spin and consequently has spin dominated by the sea. The
nucleon state vector in the chiral basis, fit to low-energy data,
gives a valence spin content consistent with experiment and lattice
QCD determinations, and has large entanglement entropy.
\end{abstract}
\maketitle

\noindent {\textbf{\textit{Introduction.}}}\ \ An important goal of
present-day nuclear science is the development of a qualitative and
quantitative understanding of the structure of the proton directly
from the underlying QCD interactions. For instance, the decomposition
of the spin of the proton into components that have a well-defined
interpretation in terms of the fundamental quark and gluon degrees of
freedom of QCD is a primary goal~\cite{Geesaman:2015fha}.  QCD reveals
that the proton is a complicated many-body quantum system, and
therefore the breakdown of its spin content is highly complex and
requires intrinsically non-perturbative methods to unravel.  Recent
work in lattice QCD addresses the nucleon spin decomposition in a
quantitative manner with controlled
uncertainties~\cite{Alexandrou:2017oeh,Lin:2018obj,Liang:2018pis}. However,
qualitative features of this decomposition, such as the suppression of
the nucleon's valence spin content, remain enigmatic, and call out for
an explanation grounded in QCD.

The complex decomposition of the nucleon spin is a striking signature
of strong entanglement among the nucleon's constituents.  How does one
characterize nucleon entanglement in a strongly-coupled quantum field
theory like QCD?  One way is to partition the nucleon state vector
into a bipartite system, trace over one of the subspaces, and obtain
an entanglement entropy. This partitioning has been done in momentum
space in the context of high-energy scattering~\cite{Kovner:2015hga},
and in position space in the context of deep-inelastic
scattering~\cite{Kharzeev:2017qzs}. These partitionings rely on being
in a regime of large momentum transfer where non-perturbative aspects
of QCD are subsumed into parton distribution functions. Many
interesting findings and potential experimental signatures are found
in these studies; a common qualitative conclusion is that in high
energy processes the nucleon constituents decohere, giving rise to a
maximal entanglement entropy given by the logarithm of the number of
gluons in the nucleon, which grows exponentially with energy in a
known manner~\cite{Kuraev:1977fs,Balitsky:1978ic}.

In this letter, it is argued that spin entanglement between valence
and non-valence spin degrees of freedom provides valuable insight into
the nucleon spin decomposition, and, more generally, into the nature
of chiral symmetry breaking in QCD. As a first step toward addressing
spin entanglement in the nucleon, special care must be taken to define
a relativistic nucleon state vector which represents the internal
degrees of freedom of the nucleon in a manner that is independent of
kinematics. This is achieved through the use of light-cone
coordinates; that is, using null-planes as quantization surfaces.  In
the Fock-space basis, which emerges naturally in light-cone
coordinates, the nucleon state vector of definite helicity is built
out of elements labeled by the number of fundamental QCD
constituents, which we will refer to in this letter as partons. The
nucleon state vector can also be expressed in a chiral basis, which
makes use of a fundamental property of the null-plane quark fields:
states of definite helicity transform irreducibly
with respect to the chiral symmetry group. The chiral basis therefore
suggests a natural bipartite Hilbert space description of the nucleon
state vector: the valence space is by definition the space which
carries non-vanishing chiral charge, while the non-valence space, or
parton sea, carries spin, but no chiral charge. The entanglement
entropy between these two subspaces drives chiral symmetry breaking in the
nucleon and provides both a qualitative and quantitative explanation
of why the valence spin content of the nucleon is so small.

\noindent {\textit{\textbf{A relativistic state vector.}}}\ \ In
non-relativistic quantum mechanics, the state vector of a many-body
system describes the internal degrees of freedom in a manner that is
independent of the choice of reference frame. This is because the
underlying kinematics is governed by the Galilean group, and Galilean
boosts do not depend on the interaction. By contrast, in a
relativistic theory of quantum mechanics, the spacetime symmetry group
is the Poincar\'e group. In general, the Poincar\'e boost operator
depends on the interaction, unless a foliation of spacetime is chosen
such that the boost operator is
non-dynamical~\cite{Dirac:1949cp}. This can be realized if null-planes
are chosen as initial quantization
surfaces~\cite{Dirac:1949cp,Kogut:1969xa,Leutwyler:1977vy}. In this
choice of light-cone coordinates, the energy and the two transverse
components of spin are dynamical, while the boosts, the momenta and
the longitudinal component of spin \textemdash the helicity
\textemdash are kinematical.  In an arbitrary reference frame, the
energy and transverse spin which act on the internal degrees of
freedom are described by the (Hamiltonian) operators $M^2$ and
$M\;\!\!\!{\cal J}_r$ ($r=1,2$), respectively, which commute with boosts
and momenta, and together with ${\cal J}_3$, satisfy the algebra of
the Poincar\'e group~\cite{Leutwyler:1977vy}.  An eigenstate of
momentum and helicity, describing a nucleon, $N$, can be expressed as
\BA
|N\, ,\,\Lambda\,;\, p^+\, ,\, {\bf p}_\perp\,\rangle = |N\, ,\,\Lambda \,\rangle \otimes |\, p^+\, ,\, {\bf p}_\perp\,\rangle \ ,
\label{eq:npstatevector}
\EA where $p^+$ and ${\bf p}_\perp$ are the longitudinal and
transverse components of the nucleon momenta, respectively\footnote{
  The conventions for the light-cone coordinates and momenta are
  defined and described in Ref.~\cite{Beane:2013ksa,Belitsky:2005qn}.}, and $\Lambda$
is the total helicity, the eigenvalue of ${\cal J}_3$.  The direct
product on the right indicates that the part of the state which
describes the internal degrees of freedom can be separated completely
from the kinematics. Achieving this separation is essential as
otherwise there is no starting point for a description of the state
vector of a nucleon which represents the internal degrees of freedom.
In principle, specification of the operators $M^2$ and $M\;\!\!{\cal J}_r$,
which act on the state $|N\, ,\,\Lambda \,\rangle$, followed by diagonalization
completely solves the dynamics. While this is intractable in QCD, the symmetry
properties of these operators can be exploited to powerful effect, as will 
be seen below. 

\noindent {\textit{\textbf{Fock-space expansion and the chiral basis.}}}\ \ The
separation of dynamics and kinematics described above allows the
Fock-basis expansion of the nucleon state, in which the state vector
is given by a (infinite) sum of contributions labeled by the QCD
field content~\footnote{Clear explanations of the QCD Fock basis are found in
  Refs.~\cite{Hornbostel:1990ya,Brodsky:2006uqa}.}.  The
contributions involving the fewest numbers of partons have been worked
out in Refs.~\cite{Ji:2002xn,Ji:2003fw,Ji:2003yj,Belitsky:2005qn}.  The dynamical
light-cone quark field $\psi_+$ is
\begin{eqnarray}
\hskip-1.3em \psi_+ (x) &=& \sum_{\lambda = \uparrow\downarrow}
\int \frac{d k^+ d^2 \bit{k}_\perp}{2 k^+ (2 \pi)^3}
 \left\{ b_\lambda (k^+ , \bit{k}_\perp) u_+(k,\lambda) {\rm e}^{- i k\cdot x}\right. \\ \nonumber
&&\left. \qquad\qquad + d^\dagger_\lambda (k^+ , \bit{k}_\perp) v_+(k,\lambda) {\rm e}^{i k\cdot x} \right\} \ ,
\label{eq:freediracmomspace}
\end{eqnarray}
where $u_+,v_+$ are Dirac wavefunctions, $b_\lambda (k^+ , \bit{k}_\perp)$ destroys a quark and
$d^\dagger_\lambda (k^+ , \bit{k}_\perp)$ creates an antiquark,
and the flavor and color indices have been suppressed. 
With standard normalization, the simplest Fock component in the proton state with $\Lambda=\oneht$ is then
\begin{eqnarray}
\label{eq:LOProtonState}
&&| u_\uparrow u_\downarrow d_\uparrow ,\oneht,{\bf 0}\rangle
\,=\,\frac{1}{2}
\int \frac{[d x] [d^2 \bit{k}_\perp]}{\sqrt{x_1 x_2 x_3}} \,
\phi (\kappa_1, \kappa_2, \kappa_3) \frac{\varepsilon^{abc}}{\sqrt{6}}\\ \nonumber
&&\times  \, u^{a \dagger}_\uparrow (x_1, \bit{k}_{1\perp})
\left\{ u^{b \dagger}_\downarrow (x_2,\bit{k}_{2\perp}) d^{\,c \dagger}_\uparrow (x_3, \bit{k}_{3\perp})
- (u\leftrightarrow d)
\right\}
| 0 \rangle
\end{eqnarray}
where $x_i$ is the longitudinal momentum of the quark in units of the
proton longitudinal momentum, the shorthand, $\kappa_i \equiv \left(
x_i , \bit{k}_{i\perp} \right)$, has been used, the integrations are
over all constituent momenta, and $\phi (\kappa_1, \kappa_2,
\kappa_3)$ is the (Fourier transform of the) light-cone wavefunction
for this particular Fock component.  Here the states are labeled by
the QCD field content, the total valence quark light-cone helicity,
and all other sources of helicity (including orbital components),
respectively. Therefore, for instance, a Fock component of the
helicity-$\oneht$ proton with, in addition to the valence quarks, one
gluon and no source of orbital angular momentum could be labeled as $|
u_\uparrow u_\downarrow d_\downarrow g_\uparrow,-\oneht,{\bf
  1}\rangle$. Note that in Eq.~\ref{eq:LOProtonState}, the proton is
in the special frame with $p^+=1$ and $\bit{p}_{\perp}=0$.  The Fock
component in a boosted frame is obtained by simply re-labeling the
momenta of the field creation operators, while leaving the
wavefunction, which carries the internal information, unchanged, in
accord with the claim made above that the internal degrees of freedom
are cleanly separated from the kinematics.  The difficulty with the
Fock expansion of the nucleon state vector is that there is no small
parameter in QCD to indicate which components should dominate out of
the infinite space of possible contributions (the large-$\nc$
approximation is an exception as we will argue below).

Assuming QCD with two massless flavors, and therefore $SU(2)_L\otimes
SU(2)_R$ chiral symmetry, the associated light-cone charges are
straightforward to construct~\cite{Beane:2013ksa}.  With these
charges, it is found that the light-cone quark fields of definite
helicity transform irreducibly with respect to the chiral group
$SU(2)_L\otimes SU(2)_R$:
\BA
\psi_{+R}\ =\ \psi_{\uparrow}\ \in \ ({\bf{ 1},\bf{ 2}}) \ \ ,  \ \
\psi_{+L}\ =\ \psi_{\downarrow}\ \in \ ({\bf{ 2},\bf{ 1}}) 
\label{eq:chiralassign}
\EA
where $({\bf {\cal R}_{L}},{\bf {\cal R}_{R}})$ labels the    
$SU(2)_L\otimes SU(2)_R$ content and ${\bf {\cal R}_{L,R}}$ are
$SU(2)_{L,R}$ representations. 

Hence in the chiral basis, the most general helicity-$\oneht$ nucleon
state, $|\,N\,,\,{\bf\oneht}\rangle$, with three valence quarks will,
in general, be a linear combination of the six classes~\footnote{The
  states with ${\bf {\cal R}_{L}}\leftrightarrow{\bf {\cal R}_{R}}$
  are contained in $|\,N\,,\,-{\bf\oneht}\rangle$.} of states: ${|
  {(\bf{2},\bf{1})}\, ,\,{\oneht}\, ,\, {\bf 0} \rangle}$, ${|
  {(\bf{2},\bf{3})_{\bf 2}}\, ,\,{\oneht}\, ,\, {\bf 0} \rangle}$, ${|
  {(\bf{1},\bf{2})}\, ,\,{-\oneht}\,,\, {\bf 1} \rangle}$, ${|
  {(\bf{3},\bf{2})_{\bf 2}}\, ,\,{-\oneht}\, ,\, {\bf 1} \rangle}$,
${| {(\bf{1},\bf{2})}\, ,\,{\threeht}\, ,\, {\bf -1} \rangle}$, ${|
  {(\bf{2},\bf{1})}\, ,\,{-\threeht}\, ,\, {\bf 2} \rangle}$, where
the states have been labeled as
\begin{eqnarray}
\!\!\!\!|\, ({\bf {\cal R}_L},{\bf {\cal R}_R})_{\bf {\cal R}}\, ,\,\lambda\, ,\, {\mathbb{p}}_3\, \rangle \ \equiv \
 |\, ({\bf {\cal R}_L},{\bf {\cal R}_R})_{\bf {\cal R}}\, ,\,\lambda \,\rangle\otimes |\, {\mathbb{p}}_3 \,\rangle \ ,
\end{eqnarray}
where ${\bf {\cal R}}$ is an $SU(2)$-isospin representation in the
product ${\bf {\cal R}_{R}}\otimes{\bf {\cal R}_{L}}$. The total helicity operator is
divided into a valence spin operator, ${\hat S}_3$, and an operator, ${\hat{\mathbb P}}_3$, that counts
all other sources of spin:
\begin{eqnarray}
  {\cal J}_3={\hat S}_3\otimes \mathbb{1}\ +\ \mathbb{1} \otimes {\hat{\mathbb P}}_3   \ ,
\end{eqnarray}
so that ${\hat S}_3|\, \lambda \,\rangle=\lambda |\, \lambda \,\rangle$,
${\hat{\mathbb P}}_3|\, {\mathbb{p}}_3 \,\rangle={\mathbb{p}}_3|\, {\mathbb{p}}_3 \,\rangle$ and
the total helicity is $\Lambda=\lambda+ {\mathbb{p}}_3$. In QCD, the gauge-invariant operator ${\hat S}_3$ is, 
up to a factor of two, given by the anomalous $U(1)_A$ chiral charge~\cite{Beane:2013ksa}. This decomposition is
therefore gauge invariant, but scale dependent.

Like the Fock basis, the chiral basis is {\it a priori} infinite dimensional, as each of
the six classes of states of definite chiral charge can
couple to any number of helicity states in the sea.  An important
advantage of the chiral basis is that the chiral transformation
properties of the energy and transverse spin are known in
QCD~\cite{Weinberg:1969hw,Weinberg:1969db,Beane:2013ksa}. In particular, both
the energy $M^2$ and the transverse spin $M\;\!\!{\cal J}_r$ transform as linear combinations of
$(\bf{1},\bf{1})$ and $({\bf 2},{\bf 2})$ representations of
$SU(2)_L\otimes SU(2)_R$; that is, in an obvious notation, ${{M}^{2}}= {{M}^{2}_{\bf{1}}} +
{{M}^{2}_{\bf{2}\bf{2}}}$ and likewise for the transverse spin.

A second, related, advantage of the chiral basis is that the
large-$\nc$ limit of the nucleon $\Lambda=\oneht$ state vector is
given by the single state ${| {(\bf{2},\bf{3})_{\bf 2}}\,
  ,\,{\oneht}\, ,\, {\bf 0} \rangle}$, whose spin is carried entirely
by the valence quarks.  The nucleon is joined in this chiral
representation by the $\Delta$ (resonance) whose $\Lambda=\oneht$
state vector is given by the single state ${| {(\bf{2},\bf{3})_{\bf
      4}}\, ,\,{\oneht}\, ,\, {\bf 0} \rangle}$ (the
$\Lambda=\threeht$ component is given by ${|{(\bf{1},\bf{4})}\,
  ,\,{\threeht}\, ,\, {\bf 0} \rangle}$). These state assignments are
easily shown to be equivalent to placing the degenerate nucleon and
$\Delta$ in the ${\bf 20}$-dimensional representation of
$SU(4)$~\cite{Weinberg:1994tu,Beane:2013ksa}, which is the large-$\nc$
expectation.  The success of the large-$\nc$ expansion in describing
nucleon properties~\cite{Gervais:1983wq,Dashen:1993as,Dashen:1993jt}
suggests that the nucleon state vector is dominated by the large-$\nc$
component with an admixture of other components which would provide
the nucleon-$\Delta$ mass splitting as well as a source of non-valence
spin.  It is unclear from the perspective of the large-$\nc$
approximation how to account for subleading corrections to the
large-$\nc$ limit in the chiral basis.  However, simple
renormalization group (RG) arguments suggest that the nucleon state
vector may be well-approximated by a small number of
components~\cite{Casher:1974xd,Susskind:1994wr}.
\begin{figure}[!t]
  \centering
     \includegraphics[scale=0.35]{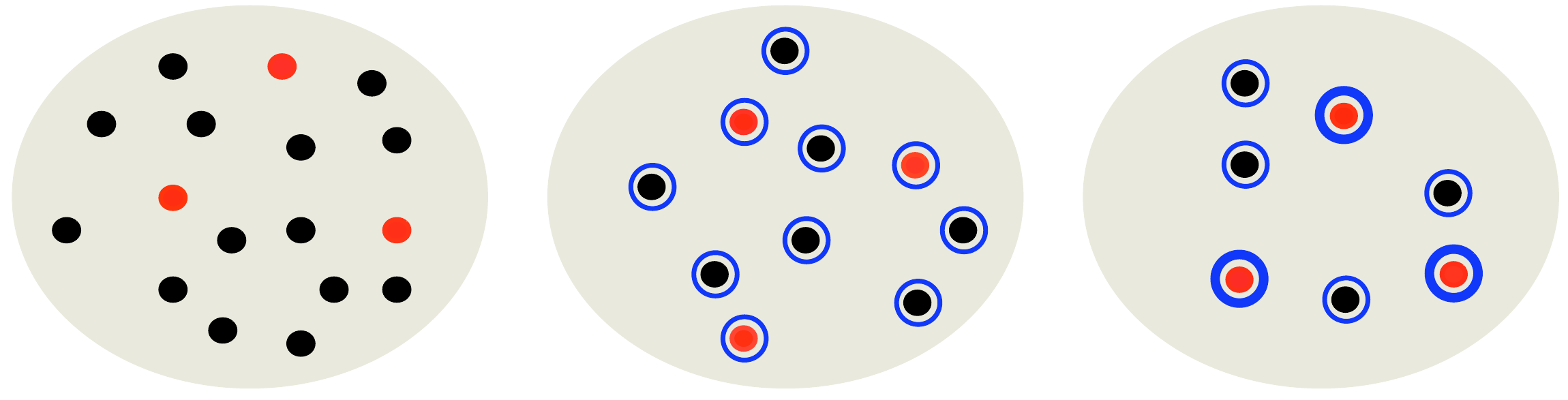}
     \caption{Nucleon as a collection of partons with some distribution in $x$ (left panel). As the partons with
low-$x$ are integrated out (middle and right panels), the ground state should be described in terms of some valence degrees of freedom (red dots) and
a finite number of sea partons (black dots), where the blue rings indicate that the partons and their interactions are dressed by the renormalization-group evolution.}
  \label{fig:rgflow}
\end{figure}

\noindent {\textit{\textbf{The renormalization group, wee partons, and the vacuum.}}}\ \ 
Consider a nucleon on the initial hyperplane defined by
$x^+=0$ and with longitudinal momentum $p^+$ and in a frame with
vanishing transverse momentum. On the initial time slice, a nucleon is defined to be a
collection of (quasi)free massless partons, each of which is labeled by $i$ and
carries a longitudinal momentum fraction ${x}_i\equiv k_i^+/p^+$
and a transverse momentum ${\bf k}_{i\perp}$ so that the nucleon
energy is
\begin{eqnarray}
M^2 &=& \sum_i\; \frac{{\bf k}^2_{i\perp}}{x_i} \ +\ \ldots
\label{eq:PMtHam}
\end{eqnarray}
In null-plane quantization small ${x}_i$ means high energy, and
therefore integrating out high-energy physics is equivalent to
integrating out low-${x}_i$ partons (so-called wee partons).
Assuming, for simplicity, that interactions are local in ${x}_i$ (nearest-neighbor),
and integrating out shells, one first integrates out all partons with
$x_i<x_{\epsilon_1}$. Integrating out these small ${x}_i$ partons result
in ``dressed'' partons that have new interactions with the ``frozen
partons'' that are represented by effective Hamiltonians with a
hierarchy of interactions governed by the small parameter
$x_{\epsilon_1}/{x}_i$~\cite{Casher:1974xd,Susskind:1994wr}. Schematically,
the energy takes the form
\begin{eqnarray}
M^2 &=& M^2_{x_i>x_{\epsilon_1}} \ +\ M^2_{x_i=x_{\epsilon_1}} \ +\ \ldots
\label{eq:PMtHamRGflow}
\end{eqnarray}
where the dots represent the frozen degrees of freedom with
$x_i<x_{\epsilon_1}$.  The first term represents the active, dynamical
partons, while the second term corresponds to the interaction of these
partons with the frozen sea. As non-perturbative physics in null-plane
quantization is carried by the low-${x}_i$ partons, this second term
carries the chiral-symmetry breaking contribution to the energy;
i.e. the part of the energy that carries non-vanishing chiral charge
and transforms as $({\bf 2},{\bf 2})$ with respect to the chiral
group: ${M}^{2}_{\bf{2}\bf{2}}$.  One can then further integrate
out the partons with $x_i<x_{\epsilon_2}$, {\it etc}. Finally this
procedure results in a description of the ground state that involves a
minimal number of partons \textemdash presumably the valence partons \textemdash
interacting with a few sea partons. This procedure of integrating out
low-${x}_i$ partons is illustrated in Fig.~\ref{fig:rgflow}.  The
resulting effective Hamiltonians are complicated, because, as noted
above, all of the vacuum physics is low-${x}_i$ physics and
therefore the RG evolution necessarily accounts
for the non-perturbative physics of QCD.  The practical consequence of
this simple RG argument is that the nucleon ground state can be described by a
nucleon state vector, with a finite number of states of definite
chiral charge, whose detailed form can be fit to experimental data. Crucially, the
truncation of the chiral basis is scale dependent as it depends on the energy cutoff, 
$x_\epsilon$.

\noindent {\textit{\textbf{Spin entanglement defined.}}}\ \ 
\begin{figure}[!t]
  \centering
     \includegraphics[scale=0.18]{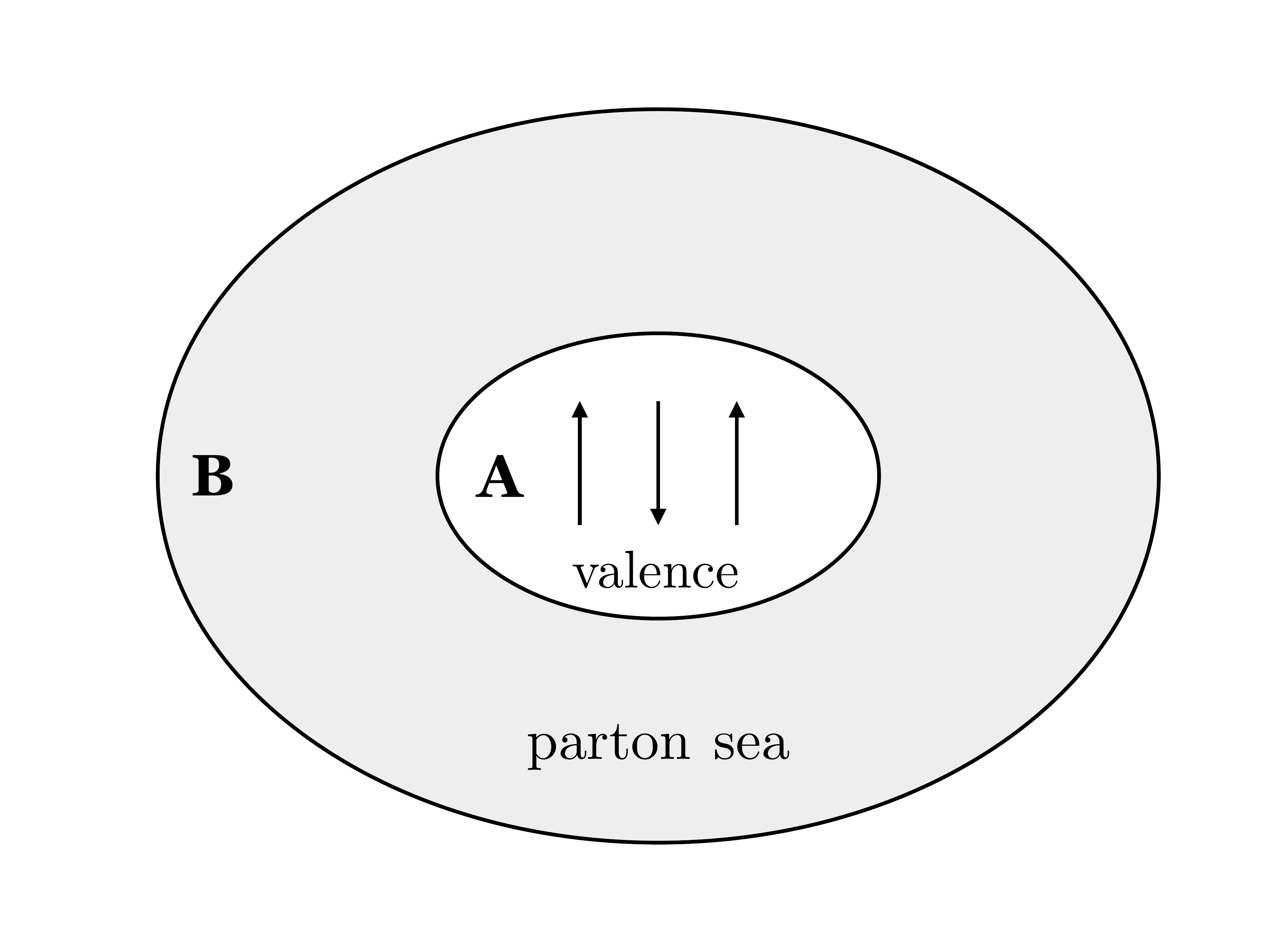}
     \caption{Bipartite Hilbert space ${\cal H}_A\otimes{\cal H}_B$ consisting of the valence quark helicity (${\cal H}_A$) and the helicity of the parton sea (${\cal H}_B$). This
     decomposition, as illustrated here, {\it does not} imply that the valence charges are in any sense localized in space.}
  \label{fig:bipartiteschema}
\end{figure}
In the chiral basis, the nucleon state vector decomposes into a
bipartite Hilbert space ${\cal H}_A\otimes{\cal H}_B$, illustrated in
Fig.~\ref{fig:bipartiteschema}, in which the valence helicity lives in
${\cal H}_A$, and all other sources of helicity live in ${\cal H}_B$. 
If measuring the valence helicity content of the nucleon, the sea is traced over to give the reduced density matrix of the valence helicity,
\begin{eqnarray}
\!\!\!\rho_A  = {\rm tr}_B\left( |\,N\,,\,{\Lambda}\,\rangle \langle \,N\,,\,{\Lambda}\,|   \right) ,
\end{eqnarray}
and the entanglement entropy is
\begin{eqnarray}
S_N = -{\rm tr}_A\left(  \rho_A\log \rho_A  \right) .
\end{eqnarray}
The general form of the nucleon state vector is
\begin{eqnarray}
\!\!\!\!\!\!\!  |N\, ,\,\Lambda \,\rangle \ =\ \sum_{i}^{n_\chi}\sum_{j}^{n_s} a_{ij}  |\, ({\bf {\cal R}_L},{\bf {\cal R}_R})_{\bf {\cal R}}\, ,\,\lambda \,\rangle_i \otimes |\, {\mathbb{p}}_3 \,\rangle_j
\label{eq:gensum}
\end{eqnarray}
where $n_\chi$ ($n_s$) is the number of valence (non-valence) states. Generally, it is expected that $n_s\gg n_\chi$, as is the case
in the Fock expansion where components with arbitrary numbers of sea partons can contribute, however, the renormalization group argument
given above suggests that many of the components are high-energy degrees of freedom that can be integrated out of the ground-state
state vector under consideration. In any event, the Schmidt decomposition theorem reveals that via a basis change, the state vector can be expressed
in a basis of dimension given by $n={\rm min} \lbrace n_\chi, n_s    \rbrace $. It is convenient to view the nucleon state vector as
a perturbation of the large-$\nc$ result, 
\begin{eqnarray}
\!\!\!  |N\, ,\,{\bf\oneht} \,\rangle \, =\, {| {(\bf{2},\bf{3})_{\bf 2}}\,,\,{\oneht}\, ,\, {\bf 0} \rangle}\, =\, {| {(\bf{2},\bf{3})_{\bf 2}}\, ,\,{\oneht} \rangle} \otimes |\, {\bf 0} \,\rangle \ ,
\end{eqnarray}
which is a product state ($n=1$) and therefore has vanishing
entanglement entropy. As the matrix element of the symmetry-breaking
energy, ${M}^{2}_{\bf{2}\bf{2}}$, vanishes between these states and
turns on only when the nucleon state vector ceases to be a
product state, it is clear that the spin entanglement defined here is
intimately related to chiral symmetry breaking and, consequently, to
the nucleon spin decomposition. Evidently, chiral symmetry is broken
in the nucleon if and only if $n>1$. Therefore, some measure of
entanglement is acting as an order parameter of chiral symmetry breaking.
This claim is considered in
generality in Ref.~\cite{Beane:2019b}. Here a simple two-component
model which captures the essence of the idea is explored.

\begin{figure}[!t]
  \centering
     \includegraphics[scale=0.33]{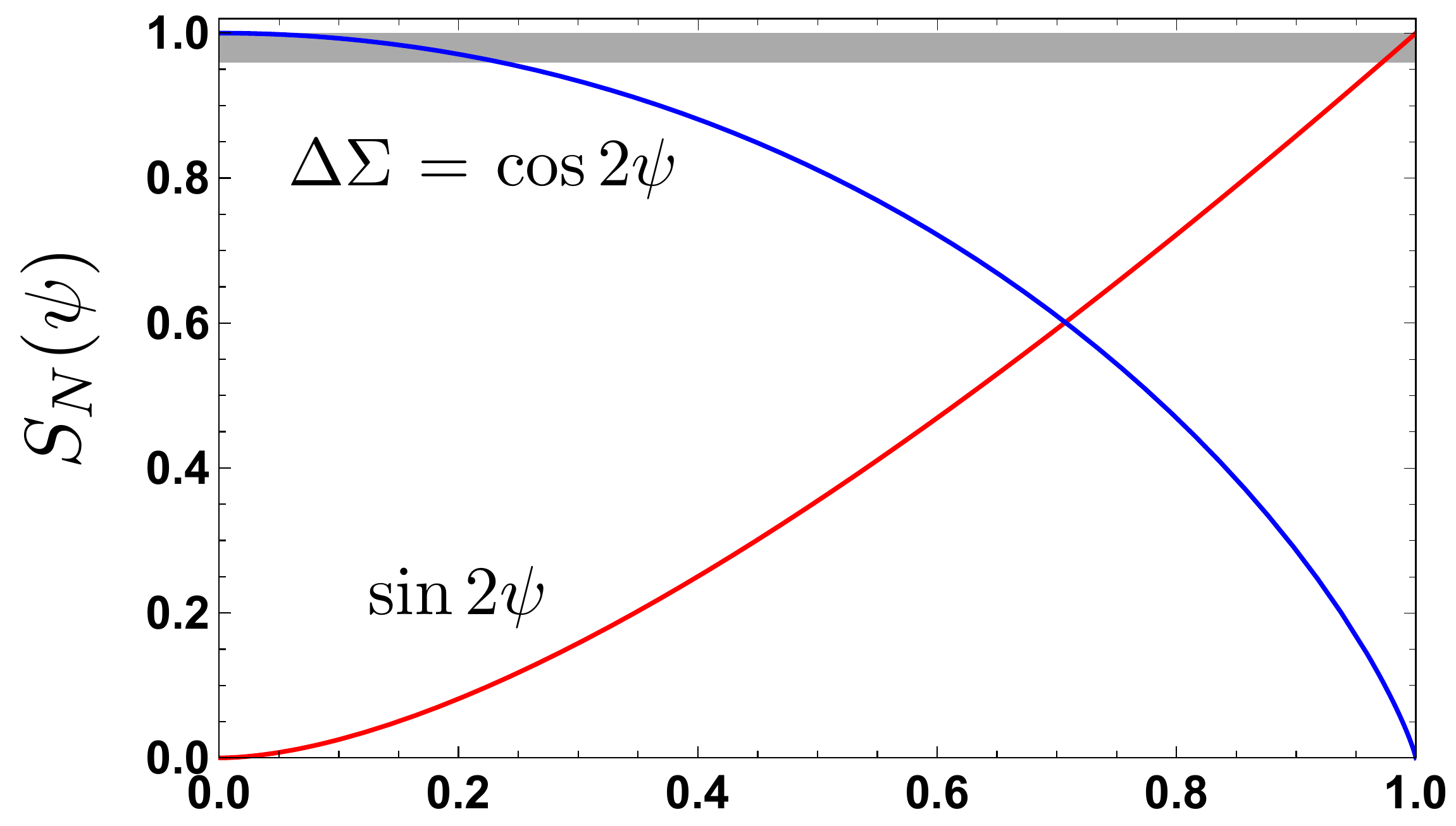}
     \caption{Entanglement entropy (in units of its maximum) versus the chiral order parameter (red curve) and the valence quark helicity content of the proton (blue curve) 
     in the minimal model of chiral-symmetry breaking. The grey band is determined by experiment.}
  \label{fig:eeminmod}
\end{figure}

\noindent {\textit{\textbf{An illustrative model.}}}\ \ 
Consider the simplest nucleon state vector that extends the large-$\nc$ result to include chiral symmetry breaking:
\begin{eqnarray}
\!\!\!\!\!\!\!\!|\,N\,,\,{\bf\oneht}\,\rangle &=& \sin\psi\,| ({\bf{1},\bf{2}}) ,{-\oneht} , {\bf 1} \rangle + 
\cos\psi\,| ({\bf{2},\bf{3}})_{\bf 2} ,{\oneht} , {\bf 0} \rangle  .
 \label{eq:spinflipmodel}
\end{eqnarray}
The state orthogonal to the nucleon in this model may be viewed as a collective of isodoublet excited states of both parities.
The vector space spans the reducible chiral representation $({\bf{1},\bf{2}})\oplus ({\bf{2},\bf{3}})$.
The large-$\nc$ limit is recovered as $\psi\rightarrow 0$, and the matrix element of the symmetry-breaking Hamiltonian scales as
\begin{eqnarray}
\langle \,N\,,\,{\bf\oneht}\,|\; M^2_{\bf 22}\; |\,N\,,\,{\bf\oneht}\,\rangle &\propto & \sin2 \psi\ .
\end{eqnarray}
The valence or intrinsic helicity contribution is
\begin{eqnarray}
\langle \,N\,,\,{\bf\oneht}\,|\; {\hat S}_3\; |\,N\,,\,{\bf\oneht}\,\rangle \ \equiv\ \oneht\Delta\Sigma \ =\ \oneht \cos 2 \psi \ .
\end{eqnarray}
Tracing over the sea gives the density matrix for the valence content
\begin{eqnarray}
\!\!\!\!\!\!\!\rho_A  &=& \sin^2\!\psi | ({\bf{1},\bf{2}})  \rangle \langle ({\bf{1},\bf{2}}) |  +\cos^2\!\psi | ({\bf{2},\bf{3}})_{\bf 2} \rangle \langle ({\bf{2},\bf{3}})_{\bf 2} | 
\end{eqnarray}
from which follows the entanglement entropy
\begin{eqnarray}
\!\!\!\!\!\!S_N(\psi)  = -\sin^2\!\psi \log\left( \sin^2\psi\right)-\cos^2\psi \log\left( \cos^2\!\psi\right) .
 \label{eq:eescheme1}
\end{eqnarray}
The probability amplitude for a spin-flip interaction is 
\begin{eqnarray}
\sin\psi &=& \langle ({\bf{1},\bf{2}})\, ,\,{-\oneht}\, ,\, {\bf 1}|\,N\,,\,{\bf\oneht}\,\rangle \ .
\end{eqnarray}
This amplitude depends not on the parton momentum fractions, which are
integrated over in the state vector, but rather on the cutoff on the
parton momentum fractions, $x_\epsilon$, as suggested by the RG arguments
given above.  A simple way of modeling this
amplitude is to neglect sea quarks and assume that there are
interactions between one of the valence quarks \textemdash which carries
longitudinal momentum fraction $x_\epsilon$, and a sea of
gluons~\cite{Carlitz:1976in}. Taking
${\cal G}(x_\epsilon)$ to be the density of the sea of gluons relative
to valence quarks and ${\cal P}(x_\epsilon)$ to be the probability of
a spin-flip interaction between the valence quark and the sea, then
$\oneht{\cal P}(x_\epsilon) {\cal G}(x_\epsilon) +1$ is the number of
spins aligned in the initial direction of the valence quark's spin and
$\oneht{\cal P}(x_\epsilon) {\cal G}(x_\epsilon)$ is the number of
spins in the opposite direction.  Assuming a purely statistical
distribution of the spins among interacting partons gives the
probability of a spin-flip interaction
\begin{eqnarray}
\sin^2\!\psi &=& \frac{\oneht{\cal P}(x_\epsilon) {\cal G}(x_\epsilon)}{1+{\cal P}(x_\epsilon) {\cal G}(x_\epsilon)}  \ .
\end{eqnarray}
Note that if there is no spin-flip interaction, then this amplitude vanishes and the large-$\nc$
result is recovered. In terms of the valence quark spin content of the nucleon,
\begin{eqnarray}
\Delta\Sigma & =& \cos 2\psi \ =\ \frac{1}{{\cal P}(x_\epsilon) {\cal G}(x_\epsilon)+1}  \ .
\end{eqnarray}
Since the density of the sea, as given by the gluon distribution function, is expected to diverge as a power law with $x_\epsilon$, it is
expected that
\begin{eqnarray}
\cos 2\psi\ \mapright{x_\epsilon\rightarrow 0}\ x_\epsilon^\delta
\end{eqnarray}
where $\delta$ is a positive number (assuming BFKL evolution for the gluon distribution function~\cite{Kuraev:1977fs,Balitsky:1978ic}, $\delta\sim 4  \alpha_s N_c \log 2/\pi$,
where $\alpha_s$ is the strong coupling constant and $N_c$ is the number of QCD colors).
As the cutoff is taken to zero and the density of gluons diverges, the nucleon is driven to the maximally entangled state
with $\psi=45^\circ$, and the nucleon spin is given entirely by the spin of the sea. Of course, this divergence is
expected to be cut off by some kind of saturation mechanism~\cite{Kovchegov:2012mbw}. The behavior of the entanglement entropy
as a function of chiral symmetry breaking and valence spin content is shown in Fig.~\ref{fig:eeminmod}.
The value of the mixing angle can be fit directly to low-energy data.
In this model, it is straightforward to find the nucleon and $\Delta$
axial and vector couplings.  For instance, $|{\cal C}_{\Delta N}|
=2\cos\psi$ and $|g_A| = \onett ( 4+\cos 2\psi )$. Fitting $|{\cal
  C}_{\Delta N}|$ to the $\Delta$-resonance decay width to pions gives
$\psi=41\pm4^\circ$ which determines, among other
things~\cite{Beane:2019b}, $g_A=1.38\pm 0.05$, $\Delta\Sigma = 0.14\pm
0.13$ and $S_N= 0.98\pm 0.02$ (in units of the maximum entropy of
$\log 2$).  This result for the valence spin content is to be compared
to $\Delta\Sigma = 0.36\pm 0.09$ taken from the JAM collaboration's
global analysis~\cite{Ethier:2017zbq}, at a renormalization scale of
$Q^2=1~{\rm GeV}^2$.

\noindent {\textit{\textbf{Realistic model-independent state
      vectors.}}}\ \ In realistic models of the nucleon state vector,
the number of states in the sum, Eq.~\ref{eq:gensum}, should grow as
the RG scale $x_\epsilon$ is reduced, with the mass of the highest
excitation in the nucleon's chiral representation setting the scale of
the separation into valence and sea spin. Detailed construction of
such realistic models is beyond the scope of this letter but may be
found in Ref.~\cite{Beane:2019b}. Many of the generic features of the
two-component model persist in more realistic state vectors: for
instance, the large-$\nc$ component is always dominant due to the
prominent nucleon-$\Delta$ axial transition, but not so much so that
it has much of an effect on the entanglement entropy, which is always
near its maximum value\footnote{By contrast, recent work has found
  that entanglement in the nuclear force, as measured by the
  entanglement power, is minimized by the experimental data, and is
  very near large-$\nc$ expectations~\cite{Beane:2018oxh}.}. The
valence spin content in these models is generally small and consistent
with the experimental range, as expected in a system where the
entanglement entropy is large.

\noindent {\textit{\textbf{Conclusion.}}}\ \ Entanglement between the
valence and non-valence helicity components of the nucleon state
vector drives chiral symmetry breaking in the nucleon. The
entanglement entropy therefore acts as an order parameter of chiral
symmetry breaking in the nucleon, and provides an explanation for why
the valence spin content of the nucleon is small in the following
sense. The interaction of the valence chiral charge with the sea,
which breaks chiral symmetry, becomes stronger as the density of sea
partons increases and the entanglement entropy rises.  The separation
between valence and non-valence spin is scale dependent, and the
renormalization group implies that a state vector with a small number
of components should suffice to provide a reasonably accurate
description of the ground state. As the highly-energetic sea partons
are integrated out, reducing the size of the vector space, the
entanglement entropy decreases and the valence spin content of the
nucleon increases.  Conversely, as more sea partons are integrated in,
increasing the size of the vector space, the entanglement entropy
increases till it reaches its maximum value where presumably some kind
of equilibrium or saturation state is achieved. It is interesting that
this pattern of RG evolution is consistent with perturbative QCD
evolution~\cite{deFlorian:2019egz}. Realistic models of
the nucleon state vector in the chiral basis which are fit to
low-energy observables give values of the valence spin content which
are consistent with experiment and with lattice QCD simulations, and
possess an entanglement entropy near its maximum value. It would be
interesting to access this chiral entanglement entropy in lattice QCD
simulations. \\

\noindent We would like to thank S.~Brodsky, A.~Deur, D.B.~Kaplan, N.~Klco, Y.~Kovchegov, W.~Melnitchouk, M.J.~Savage and R.~Venugopalan
for valuable communications. This work was supported in part by the
U. S. Department of Energy grant DE-SC001347.  


\bibliography{/Users/silasbeane/silas/Physics_Projects/FrontForm/SpinContent/bibi}

\end{document}